\documentclass[aps,prl,twocolumn,tightenlines,showkeys,showpacs,floatfix,lengthcheck]{revtex4}
\usepackage[dvips]{color,graphicx}

\def\Journal#1#2#3#4{{#1} {\bf #2}, #3 (#4)}
\def\EPJA{{\rm Eur. Phys. J.} A}
\def\EPJD{{\rm Eur. Phys. J.} D}

\def\NPA{{\rm Nucl. Phys.} A}

\def\PR{\rm Phys. Rev.}
\def\PRL{\rm Phys. Rev. Lett.}
\def\PRD{{\rm Phys. Rev.} D}
\def\PRC{{\rm Phys. Rev.} C}

\def\be{\begin{eqnarray}}
\def\ee{\end{eqnarray}}
\def\bea{\begin{eqnarray}}
\def\eea{\end{eqnarray}}
\def\beas{\begin{eqnarray*}}
\def\eeas{\end{eqnarray*}}
\newcommand{\eq}[1]{Eq.~(\ref{#1})}

\def\bfb{{\bf b}}

\newcommand{\boldmu}{\mbox{\boldmath $\mu$}}

\def\bfq {{\bf q}}

\def\bfp{{\bf p}}
\begin{document}

\noindent
{ NT@UW-07-16}
\title{
 Proton  Electromagnetic Form Factor  Ratios at Low Q$^2$
}
\author{ 
Gerald   A. Miller}

\affiliation{Department of Physics,
University of Washington\\
Seattle, Washington 98195-1560
}

\author{E. Piasetzky and G. Ron}
\affiliation{The Beverly and Raymond Sackler School of Exact Sciences\\ Tel Aviv University\\
Tel Aviv 69978, Israel}

\begin{abstract}
We study 
the ratio $R\equiv\mu G_E(Q^2)/G_M(Q^2)$ of the
proton at very small values of $Q^2$. 
Radii commonly associated with these form 
factors are not moments of charge or magnetization densities. We show that the  form factor $F_2$ is  
correctly interpretable as the two-dimensional Fourier transformation of a magnetization
density. A relationship between the measurable ratio and moments of true charge and magnetization 
densities is derived. We find that existing measurements 
show that the magnetization 
density  extends  further than the charge density, in contrast with expectations
based on the measured reduction of $R$ as $Q^2$ increases.


\end{abstract}


\maketitle

Electromagnetic form factors of the proton and neutron (nucleon) 
are  probability amplitudes that the  nucleon can absorb a given amount 
of momentum and  remain in the ground state, and therefore  should determine
the nucleon charge and magnetization densities. Much experimental technique, 
effort and ingenuity has been used recently to measure these quantities~\cite{reviews,chw}.

The  text-book  interpretation of electromagnetic  form factors, $G_E,G_M$,
 explained in \cite{chw},
is  that their Fourier transforms are measurements of charge and magnetization densities, and
conventional wisdom relates the charge and magnetization mean 
square  radii to the 
slopes of 
$G_{E,M}$ at $Q^2=0$.  
However, this interpretation is not correct because  the   wave functions of the
initial and final  nucleons have different momenta and therefore differ, 
 invalidating a probability or density
interpretation. A proper charge density is   related to the matrix element 
of an absolute square of a field operator.

Here we show that the proper magnetization  density is
the two-dimensional Fourier transform  of the $F_2$ form factor. 
We  use this  and the  result that the charge density 
is the two-dimensional Fourier transform  of the $F_1$ form factor
\cite{soper1,mbimpact,diehl2,myneutron} to show that 
 the magnetization density of the proton extends significantly
further  than its charge density. This is surprising because the observed rapid 
decrease of the ratio of electric to magnetic form factors
with increasing  values of  $Q^2$ ~\cite{reviews,chw} might lead one
to conclude that the charge radius is larger  than the magnetization radius.

Form factors are 
matrix elements of the electromagnetic current operator 
 $J^\mu(x^\nu)$ in units of the proton 
charge:
\begin{widetext}  
\bea
\langle p',\lambda'| J^\mu(0)| p,\lambda\rangle
=\bar{u}(p',\lambda')\left(\gamma^\mu F_1(Q^2)+i{\sigma^{\mu\alpha}\over 2M}q_\alpha F_2(Q^2)
\right) u(p,\lambda),\eea\end{widetext}
where the momentum transfer $q_\alpha=p'_\alpha-p_\alpha$ is taken as space-like, 
so that $Q^2\equiv -q^2>0.$ The nucleon polarization states are  those of 
definite light-cone helicities $\lambda,\lambda'$ \cite{soper}.  
The normalization is such that  $F_1(0)$ is the
nucleon charge, and  $F_2(0)$ is the proton
anomalous magnetic moment $\kappa=1.79$. The Sachs form factors 
are
$G_E(Q^2)\equiv 
F_1(Q^2)-{Q^2\over 4M^2}F_2(Q^2),\; G_M(Q^2)\equiv F_1(Q^2)+F_2(Q^2).$ 
Any probability or density 
interpretation of $G_E$ is spoiled by a non-zero  value of $Q^2$, no matter how small. 
Nevertheless, $G_E,G_M$ are experimentally accessible so we 
define {\it effective (*)} square radii $R_E^{*2},R_M^{*2}$
 such that for small values of $Q^2$: 
$G_E(Q^2)\approx 1-\frac{Q^2}{6}R_E^{*2},\; 
G_M(Q^2)\approx\mu( 1-\frac{Q^2}{6}R_M^{*2})$, 
where for the proton $\mu=2.79$. Thus the accurately measurable \cite{newprop}
 ratio
\bea 
\mu G_E(Q^2)/G_M(Q^2)\approx 1+\frac{Q^2}{6}\; (R_M^{*2}-R_E^{*2}).
\label{def2}\eea 

The form factor $F_{1}$ is a 
 two-dimensional Fourier transform of the true charge density $\rho(b)$, where
 $b$ is the distance 
from the transverse center of mass position irrespective of the
longitudinal momentum
\cite{soper1}-\cite{myneutron}, with
\bea
F_1(Q^2=\bfq^2)=\int d^2b
\;\rho(b)e^{-i\bfq\cdot\bfb}.
\label{rhob}\eea
At small values of  $\bfq^2$, 
\bea
F_1(Q^2=\bfq^2)&\approx&
1-\frac{Q^2}{4}\langle b^2\rangle_{Ch}
\label{rhobexp}\eea
where $\langle b^2\rangle_{Ch}$ 
is the second moment of  $\rho(b)$.

We now derive a similar interpretation for $F_2$ in terms of a magnetization density, 
starting with  the relation that $\boldmu\cdot{\bf B}$ is the matrix element of
${\bf J}\cdot{\bf A}$ in a definite state, $\vert X\rangle$. 
Take the rest-frame magnetic field
to be a constant vector in the $1$ (or   $b_x$) \cite{notation}
  direction, and the corresponding vector
potential as 
${\bf A}= B b_y \hat{{\bf z}}$. Then  consider the
system in a 
frame in which the plus component of the momentum
approaches infinity. The anomalous magnetic
moment may be extracted 
by taking:
\bea \vert X\rangle\equiv \frac{1}{\sqrt{2}}\left[\left|p^+,{\bf R}= {\bf 0},
+\right\rangle+\left|p^+,{\bf R}= {\bf 0},
-\right\rangle\right],\eea
where  $\left|p^+,{\bf R}= {\bf 0},+\right\rangle $ represents a
transversely localized state of definite $P^+$ and light-cone helicity. 
The state $\vert X\rangle$ \cite{mbprd66,mbimpact}
may be interpreted as that of
  a transversely polarized target, up to relativistic corrections caused by
the transverse localization of the wave packet \cite{mbprd72}. The anomalous 
magnetic moment $\mu_a$ \cite{anom} is then given by 
\bea \mu_a=
\frac{\langle X\vert \int dx^-d^2b\; b_y\;\bar{q}(x^-,\bfb)\gamma^+q(x^-,\bfb)\vert X\rangle}
{\langle X\vert X\rangle}.\eea Use  translational invariance to obtain: 
\bea \mu_a=
{\langle X\vert\int d^2b \;b_y\;{q}_+^\dagger(0,\bfb)q_+(0,\bfb)\vert X\rangle},\label{mua}\eea
where  $q(x^-,\bfb)$ is a quark-field operator, and $q_+=\gamma^0\gamma^+q$.
This  matrix element of a quark density 
operator 
is closely related to the Burkardt's \cite{mbprd66,mbimpact} 
impact parameter GPD $q_X(x=0)$:
\begin{widetext}
\bea 
q_X(x,{\bf b})\equiv \langle X\vert \int \frac{dx^-}{4\pi}{q}_+^\dagger
\left(0 
,{\bf b} \right) 
q_+\left({x^-},{\bf b}\right) 
e^{ixp^+x^-}\vert X\rangle
={1\over 2p^+}( {\cal H}_q(x,\xi=0,b)-{1\over 2 M}\frac{\partial}{\partial b_y}{\cal E}_q(x,\xi=0,b)),
\label{mb} \eea
\end{widetext}
where $ {\cal H}_q$ and  $ {\cal E}_q$ are two-dimensional Fourier transforms
of the GPDs $H_q,E_q$ \cite{gpda}.
Integration of \eq{mb} over $x$ sets $x^-$  to zero, so that
\eq{mua} can be re-expressed (after integration by parts) as
 \bea
\mu_a={1\over 2p^+}\int d^2b \int dx\; {\cal E}_q(x,\xi=0,b).\label{muax}\eea
But the integral of ${\cal E}_q$ over $x$ is just the two-dimensional 
Fourier transform of  $2p^+F_2$, so that 
\bea \mu_a=\int d^2b \;\rho_M(b),   \eea
where
\bea
\rho_M(b)=\int\frac{d^2q}{(2\pi)^2} F_2(t=-\bfq^2)e^{i \bfq\cdot \bfb}.\label{mdensx}\eea
The subscript $M$ denotes that this density generates the 
anomalous magnetic moment, is properly  a true magnetization density,
 and is distinct from  $\rho(b)$. It is also possible to consider
the quantity $-\int dx b_y{\partial\over \partial b_y}{\cal E}_q(x,0,b)$ as
the magnetization density. However, this definition would depend on the choice
of the $x$ axis as the  direction of the magnetic field. A true intrinsic
quantity should not depend on such  a choice, so we use the 
form of Eqs.~(\ref{muax}-\ref{mdensx}).

For  small values of $\bfq^2$:
\bea
F_2(Q^2=\bfq^2) \approx\kappa\left(1-\frac{Q^2}{4}\langle b^2\rangle_{M}\right),
\label{rhobt}\eea
where $\langle b^2\rangle_{M}$ is the second moment of 
$\rho_M(b)$. 
Use the definitions of the Sachs form factors, \eq{def2}
and  the expansions Eqs.~(\ref{rhobexp},\ref{rhobt})
to relate the true moments to the effective square radii so that 
\bea
\langle b^2\rangle_{M}-\langle b^2\rangle_{Ch}={\mu\over \kappa}{2\over3}(R_M^{*2}-R_E^{*2})+
{\mu\over M^2}
.\label{formula}\eea
The low $Q^2$ measurement of the form factor ratio determines also
the difference of true moments $\langle b^2\rangle_{M}-\langle b^2\rangle_{Ch}$ 
which is approximately
 the difference of the effective square radii plus
a specific relativistic correction ${\mu\over M^2}\approx0.1235 \;{\rm fm}^2.$
This  
is the consequence of the Foldy term \cite{foldy},
arising from the interaction of the anomalous magnetic moment of the nucleon with
the external magnetic field of the  electron. 

\begin{center}
 \begin{figure}
   \centering
   \includegraphics[width=3.5in]{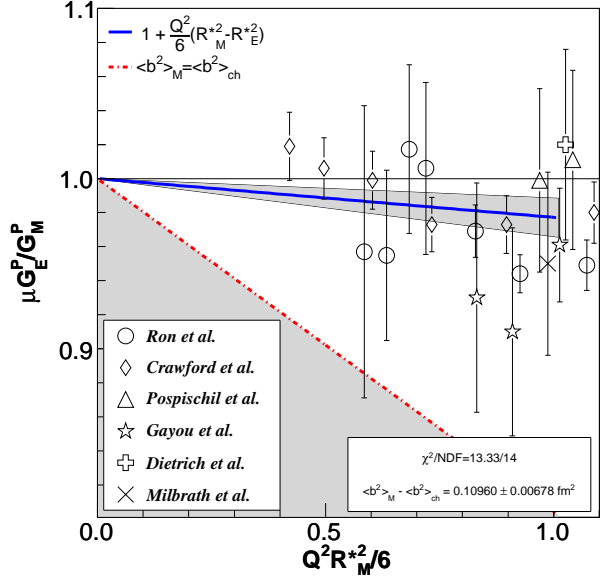}  
   \caption{[Color Online] Results of a linear fit to world dataset of high 
     precision polarization transfer 
     experiments~\cite{ron2007,cra2007,pop2001,gayou01,die2001,mil1998}, shown by the
     solid (blue) line and error  band. The shaded area indicates 
     $\langle b^2\rangle_{Ch}>\langle b^2\rangle_M$. 
The dashed  (red) line shows the 
     critical slope  $S_c=-\frac{3}{2}\frac{\kappa}{M^2}$ (\eq{formula})   
     giving $\langle b^2\rangle_M=\langle b^2\rangle_{Ch}$.}
     \label{fig:fit}
 \end{figure}
\end{center}

Using the high precision results from polarization transfer 
experiments (generally accepted to be less
sensitive to two photon exchange effects~\cite{arr2004}) 
we compare the world's data set for $\mu G_E/G_M$ with Eq.~(\ref{def2}) in
Fig.~\ref{fig:fit}.
We fix the value of $R_M^*$  from  a new state-of-art determination~\cite{volotka},
$R_M^*=0.778(29)$ fm and plot the data as a function of  the 
 parameter
$\frac{Q^2}{6}{R_M^*}^2$. 
We find
\bea 
\left<b^2\right>_M-\left<b^2\right>_{ch} = 0.10960\pm0.00678\;{\rm fm}^2,
\label{result}
\eea
which is about 12\% smaller than the contribution of the magnetic moment  Foldy term.
Thus the difference 
${2\over 3}{\mu\over\kappa}\left(R^{*2}_M-R^{*2}_E\right)=-0.0139\pm0.00678 $ fm$^2$
presently  has the opposite sign of the result for the difference of the true 
moments of the distribution,
indicating  the 
need to base interpretations on the true moments. Note also that
$\left(R^{*2}_M-R^{*2}_E\right)$ is determined to an accuracy of only about 50\%.
Improving the accuracy 
can only  be achieved by  using  very small values of $Q^2$, for which 
no high precision polarization
 transfer results exist.
Two-photon effect corrected cross
 section measurements of the proton form factor ratio at very low $Q^2$,
 however, are consistent with the ratio $R=1$ \cite{arrcfe2006},  
corresponding to $R_M*=R_E*$.
 in rough agreement with our results.

\begin{center}
 \begin{figure}
   \centering
   \includegraphics[angle=0,width=3.5in]{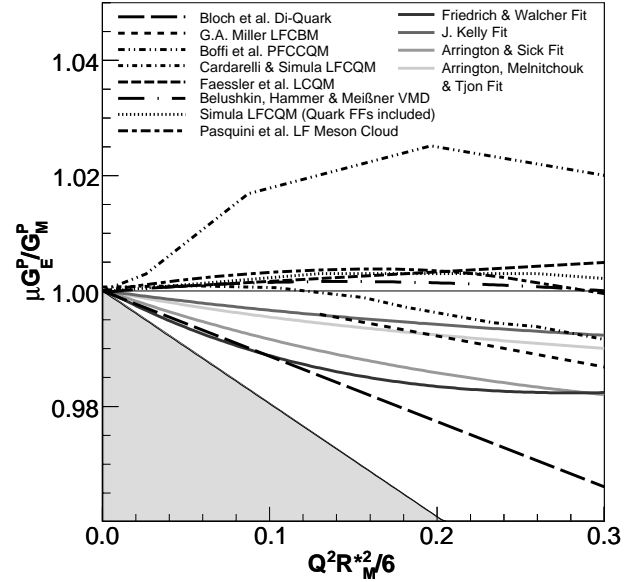}
   \caption{The proton EM form factor ratio from several recent calculation and fits. The 
     shaded area indicates $\left<b^2\right>_{Ch}>\left<b^2\right>_M$. The
     fits are shown using  solid black curves with different line types, and 
     the calculations as solid lines using different shades of gray.}
   \label{fig:calcs}
 \end{figure}
\end{center}

\begin{center}
 \begin{figure}[ht]
   \centering
   \includegraphics[width=3.5in]{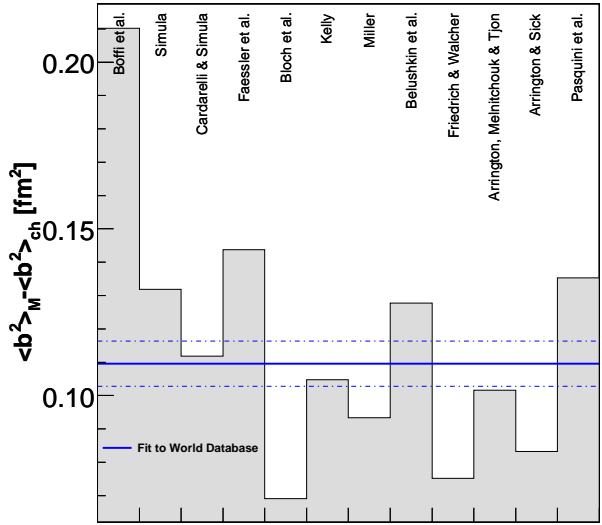}
   \caption{[Color online] $\left<b^2\right>_{M}-\left<b^2\right>_{Ch}$ from 
     recent calculations and fits. All fits and calculations yield a positive value.
     The solid (blue) line shows the result \eq{result}. } 
   \label{fig:allrange}
 \end{figure}
\end{center}

Figure~\ref{fig:calcs} shows several model calculations and 
fits for the form factor 
ratio~\cite{mil2002,belushkin2006,fandw2003,arr2004,arrcfe2006,simula-2001,card2000,faessler06,boffi2002,Bloch:1999ke,kel2004,Pasquini:2007iz}, which are seen to vary greatly.
 Improved experiments \cite{newprop} 
would be able to distinguish  these diverse approaches, and more fundamentally, 
 better determine the value of
$\left(R^{*2}_M-R^{*2}_E\right)$.
We  also use a linear fit, at small values of  $Q^2$, 
to the results of various calculations and some global fits. 
 These 
are  shown  in Fig.~\ref{fig:allrange}. While there is significant  variation, 
all 
agree  with our  result  \eq{result}.

Our   result that the magnetization density extends further than the
charge density is consistent with the failure of the   spin of 
the quarks to account for 
the angular momentum  of the proton \cite{oldspin}, and the likley importance 
of quark orbital angular momentum (OAM).
This is because quarks carrying OAM, and therefore much magnetization-generating
current,   
are located away from the center.
For example, 
consider
the pion  cloud, which dominates the proton's exterior, 
as a  source of OAM.
The pion cloud is  more influential 
for magnetic properties than for electric ones
({\it e.g} Refs.~\cite{cbm,Hammer:2003qv}),  and causes 
a  proton magnetization radius that is 
 larger than the charge radius. 

To reiterate: our model independent result is that 
the magnetization density of the proton extends further than
its charge density.  A natural interpretation 
involves  the
orbital angular momentum  carried by quarks. Future experimental measurements
of the ratio of the proton's   electromagnetic form factors would render the present results more precise.

We thank the USDOE,  (FG02-97ER41014), 
the Israel Science Foundation, and the Adams Fellowship Program of the
Israel Academy of Sciences and Humanities for partial support of this work. 
We thank  M.~Burkardt, J.~Arrington, R.~Gilman, T.~Goldman, and 
 D.~Higinbotham
 for useful discussions.


\begin{thebibliography}{0}

\bibitem{reviews} 
 H.~y.~Gao,
  Int.\ J.\ Mod.\ Phys.\  E {\bf 12}, 1 (2003)
  [Erratum-ibid.\  E {\bf 12}, 567 (2003)];
  C.~F.~Perdrisat, V.~Punjabi and M.~Vanderhaeghen,
Prog.\ Part.\ Nucl.\ Phys.\  {\bf 59}, 694 (2007).
 J.~Arrington, C.~D.~Roberts and J.~M.~Zanotti,
  J.\ Phys.\ G {\bf 34}, S23 (2007).
\bibitem{chw} C.~E.~Hyde-Wright and K.~de Jager,
  Ann.\ Rev.\ Nucl.\ Part.\ Sci.\  {\bf 54}, 217 (2004);

\bibitem{soper1} D.E. Soper, Phys.\ Rev.\ D\ {\bf 15},
 1141 (1977). 

\bibitem{mbimpact}
  M.~Burkardt,
  Int.\ J.\ Mod.\ Phys.\  A {\bf 18}, 173 (2003).

\bibitem{diehl2} 
  M.~Diehl,
  Eur.\ Phys.\ J.\  C {\bf 25}, 223 (2002)
  [Erratum-ibid.\  C {\bf 31}, 277 (2003)].

\bibitem{myneutron} G.~A.~Miller,
  Phys. Rev. Lett. {\bf 99}, 112001 (2007)
  
\bibitem{soper} J. Kogut and D.E. Soper, 
  Phys.\ Rev.\ D\ {\bf 1}, 2901 (1970).



\bibitem{newprop} G.~Ron {\it et. al.}, {J}efferson Lab Proposal PR07-004.

\bibitem{notation}
  Our notation is:
 $x^\pm\equiv (x^0\pm x^3)/\sqrt{2},p^\pm\equiv (p^0\pm p^3)/\sqrt{2}$, and
  $p_\mu x^\mu=p^-x^++p^+x^--\bfp\cdot\bfb$. 


\bibitem{mbprd66}
  M.~Burkardt,
  Phys.\ Rev.\  D {\bf 66}, 114005 (2002).


\bibitem{mbprd72}
  M.~Burkardt,
  Phys.\ Rev.\  D {\bf 72}, 094020 (2005).

\bibitem{anom}
  The  matrix element in the  
transversely localized state $\vert X\rangle$,
  suppresses the Dirac contribution 
  \cite{mbprd72}, leaving only the effects of the anomalous term. 
\bibitem{gpda}
  D. M\"uller et al., Fortschr. Phys. {\bf 42}, 101 (1994);
  X.Ji, J.\ Phys.\ G\ {\bf 24}, 1181 (1998);
  A.V. Radyushkin, Phys.\ Rev.\ D\ {\bf 56}, 5524 (1997).
\bibitem{ron2007}G.~Ron {\it et. al.} , Phys. Rev. Lett. {\bf 99}, 202002 (2007).
\bibitem{cra2007}C.B.~Crawford {\it et. al.}, \Journal{\PRL}{98}{052301}{2007}.
\bibitem{pop2001}T.~Pospichil {\it et. al.}, \Journal{\EPJA}{12}{125}{2001}.
\bibitem{gayou01}O.~Gayou {\it et. al.}, \Journal{\PRC}{64}{038202}{2001}.
\bibitem{die2001}S.~Dietrich {\it et. al.}, \Journal{\NPA}{690}{231}{2001}.
\bibitem{mil1998}B.D.~Milbrath {\it et. al.}, \Journal{\PRL}{80}{452}{1998}.
\bibitem {foldy} L.~L.~Foldy, Phys.\ Rev.\ {\bf83}, 688  (1951).
\bibitem{arr2004}J.~Arrington, \Journal{\PRC}{69} {022201} {2004}
\bibitem{volotka}A.V.~Volotka {\it et. al.}, \Journal{\EPJD}{33}{23}{2005}.

\bibitem{mil2002}M.~R.~Frank, B.~K.~Jennings and G.~A.~Miller,
  Phys.\ Rev.\  C {\bf 54}, 920 (1996);
  G.A.~Miller, \Journal{\PR}{66}{032201(R)}{2002}.
\bibitem{belushkin2006}M.A.~Belushkin {\it et. al.}, \Journal{\PRC}{75}{035202}{2007}.
\bibitem{fandw2003}J.~Friedrich and T.~Walcher, \Journal{\EPJA}{17}{607}{2003}.
\bibitem{arrcfe2006}J.~Arrington {\it et. al.}, \Journal{\PRC}{76}{035201}{2007}.
\bibitem{simula-2001}S.~Simula, Proceedings of the Int'l Conf. on "The Physics of Excited Nucleons" (N$^*$ 2001).
\bibitem{card2000}F.~Cardarelli {\it et. al.}, \Journal{\PRC}{62}{065201}{2000}.
\bibitem{faessler06}A.~Faessler {\it et. al.}, \Journal{\PRD}{73}{114021}{2006}.
\bibitem{boffi2002}S.~Boffi {\it et. al.}, \Journal{\EPJA}{14}{17}{2002}.
\bibitem{Bloch:1999ke}J.C.R.~Bloch {\it et. al.}, \Journal{\PRC}{60}{062201}{1999}.
\bibitem{kel2004}J.J.~Kelly, \Journal{\PRC}{70}{068202}{2004}.
\bibitem{Pasquini:2007iz}
  B.~Pasquini and S.~Boffi,
  Phys.\ Rev.\  D {\bf 76}, 074011 (2007).
\bibitem{oldspin}  E.~W.~Hughes and R.~Voss,
  ARNPS,  {\bf 49}, 303 (1999).
\bibitem{cbm}  A.~W.~Thomas, S.~Th\'eberge and G.~A.~Miller,
  Phys.\ Rev.\  D {\bf 24}, 216 (1981).
\bibitem{Hammer:2003qv}
  H.~W.~Hammer {\it et. al.}, 
  Phys.\ Lett.\  B {\bf 586}, 291 (2004).

\end{thebibliography}
\end{document}